\documentclass[showpacs, pre]{revtex4}   

\usepackage{amsmath}
\usepackage{amssymb}
\usepackage{times}
\usepackage{verbatim}
\usepackage[applemac]{inputenc}
\usepackage[T1]{fontenc}
\usepackage{lmodern}
\usepackage{wrapfig}
\usepackage{graphicx}
\usepackage{rotating}
\usepackage{multirow}
\usepackage{float}

\DeclareGraphicsExtensions{.eps}

\begin{document}

\title{Supercooperation in Evolutionary Games on Correlated Weighted Networks}
\author{Pierre Buesser}
\email{pierre.buesser@unil.ch}
\affiliation{Information Systems Institute, HEC, University of Lausanne, Switzerland}

\author{Marco Tomassini}
\email{marco.tomassini@unil.ch}
\affiliation{Information Systems Institute, HEC, University of Lausanne, Switzerland}

\begin{abstract}
In this work we study the behavior of classical two-person, two-stra\-te\-gies evolutionary games
on a class of weighted networks derived from Barab\'asi-Albert and random scale-free unweighted
graphs. Using customary imitative dynamics, our numerical simulation results show that the
presence of link weights that are correlated in a particular manner with the degree of the link endpoints,
leads to unprecedented levels of cooperation in the whole games' phase space, well above
those found for the corresponding unweighted complex networks. We provide
intuitive explanations for this favorable
behavior by transforming the weighted networks into unweighted ones with particular topological properties.
The resulting structures help to understand why cooperation can thrive and also give
ideas as to how such supercooperative networks might be built.
\end{abstract}

\pacs{89.75.Hc, 87.23.Ge, 02.50.Le, 87.23.Kg}

\maketitle

\section{Introduction}
\label{abstract}

Game theory has proved useful in a number of settings in biology, economy, and social science in general. Evolutionary
game theory in particular is well suited to the study of strategic interactions in animal and human populations and
has been a very useful mathematical tool for dealing with these kind of situations.
Evolutionary games have been traditionally studied in the context of well-mixed and very large populations 
(see e.g.~\cite{weibull95,Hofbauer1998}). However, starting with the work of Nowak and May~\cite{nowakmay92}, and
especially in the last few years, population structures with local interactions have been brought to the focus
of research. Indeed, it is a fact of life that social interactions can be more precisely
represented as networks of contacts in which nodes represent agents and links stand for their 
relationships~\cite{newman-book}.
The corresponding literature has already grown to a point that makes it difficult to be exhaustive; however,
good recent reviews can be found in~\cite{szabo,anxo1,mini-rev}.
 Most of the new results, owing to the difficulty of analytically solving non mean-field models, come from numerical simulations, 
 but there are also some theoretical results, mainly on degree-homogeneous graphs.
On the other hand, most of the work  so far has dealt with unweighted graphs.
This is understandable as a logical first step, since attributing reliable weights to relationships in social networks is not a simple matter because
the relationship is often multi-faceted and  implies psychological and sociological features
that are difficult to define and measure, such as friendship, empathy, and common beliefs. 
In spite of these difficulties, including the strength of agents' ties would be a step toward more realistic models. In fact, sometimes
at least a proxy for the intensity of a relationship can be defined and accurately measured. This is the case for e-mail networks, phone calls
networks, and coauthorship networks among others. For example, in an extensive study of mobile phone calls 
network~\cite{mobile-nets2}, the authors used
the number of calls between two given agents and the calls' duration to capture at least part of the underlying more complex social interaction.
The same can be done in coauthorship networks in which the strength of a tie can be related to the number of common
papers written by the two authors~\cite{newman-collab-2}. 
\begin{figure}[ht]
\begin{center}
 \includegraphics[width=7cm]{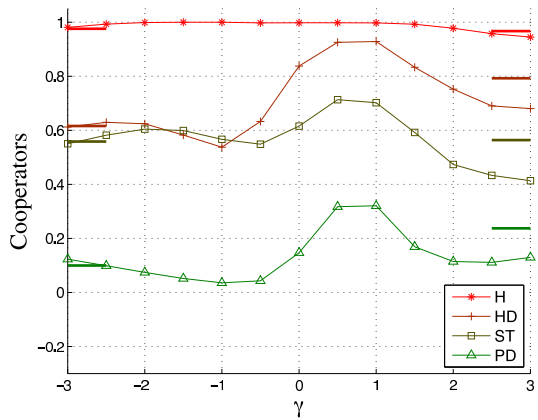}
 \caption{ (Color online) The points show, for each game, the average fraction of cooperation at steady state in the whole
 game's phase space as a
 function of $\gamma$, $-3 \le \gamma \le 3$ in Barab\'asi--Albert networks. The weight-degree correlation is $w_{ij}=(k_ik_j)^\gamma$ 
 and the strategy update rule is replicator dynamics. 
 Each point is the average of $50$ independent runs; continuous curves are just a guide for the eye.
 Small lines on the right and left borders indicate the cooperation level for $\gamma= \pm 10$.}
\label{cadran}
\end{center}
\end{figure}

Little work has been done on evolutionary games on weighted networks to date. Two recent contributions are ~\cite{chinois08,VoelklBL09}.
We have recently offered a rather systematic analysis of standard evolutionary games on several network types
using some common weight distributions with no correlations with the underlying network topology features such as node degree and
local clustering~\cite{weighted-physicaA}.
The results in all cases studied in~\cite{weighted-physicaA} are that the evolutionary game dynamics are little affected
by the presence of weights, that is, the graph topology, together with the game class and the strategy update rules seem to be the main 
factors dictating strategy evolution in the population.
This is a reassuring result to the extent to which it allows us to almost ignore the weights and focus on the structural aspect only.
However, the previous results, as said above, were obtained with standard weight distribution functions and ignoring possible
correlations with other topological features. Thus, it is still possible that some particular weight assignment that takes these factors
into account could make 
the dynamics to behave in a more radically different way. Indeed, in~\cite{chinois08}, Du et al. have investigated
the Prisoner's Dilemma on Barab\'asi--Albert scale-free graphs with a particular form of two-body correlation between the link weights 
and the degrees of the link endpoints, finding an increase of cooperation under some conditions with 
respect to the unweighted case. We shall discuss and extend 
their results in Sect.~\ref{deg-weight-corr}.
 In the present paper we have studied another form of degree-weight
correlation that is more likely to occur in social networks and gives unprecedented amounts of cooperation promotion on scale-free
networks. In the following we first give a brief description of the games used and of the population dynamics; then we present 
and discuss our results.

\section{Evolutionary Games on Networks}

\subsection{The Games Studied}

We investigate four classical two-person, two-strategy, symmetric games, namely the Prisoner's
Dilemma (PD),
the Hawk-Dove Game (HD), the Stag Hunt (ST), and the Harmony game (H). We  briefly summarize
the main features of these games here for completeness; more detailed accounts
can be found elsewhere~\cite{weibull95}.
The games have the generic payoff bi-matrix of Table~\ref{pbm}.

\begin{table}[t]
\begin{center}
{\normalsize
$
\begin{array}{c|cc}
 & C & D\\
\hline
C & R,R & S,T\\
D & T,S & P,P
\end{array}
$}
\end{center}
\caption{Generic payoff bi-matrix for the two-person, two-strategies symmetric games
discussed in the text. \label{pbm}}
\end{table}
\noindent The set of strategies is $\Lambda=\{C,D\}$, where $C$ stands for ``cooperation'' and $D$ means ``defection''.
In the payoff matrix $R$ stands for the \textit{reward}
the two players receive if they
both cooperate, $P$ is the \textit{punishment} if they both defect, and $T$  is the
\textit{temptation}, i.e.~the payoff that a player receives if he defects while the
other cooperates getting the \textit{sucker's payoff} $S$.
In order to study the standard parameter space, we restrict the payoff values in the following
way: $R=1$, $P=0$, $-1 \leq S \leq 1$, and $0 \leq T \leq 2$. In the resulting $TS$-plane, each game corresponds
to a different quadrant depending on the ordering of the payoffs.

\noindent For the PD, the payoff values are ordered such that $T > R > P > S$. 
Defection is always the best rational individual choice, so that 
$(D,D)$ is the unique Nash Equilibrium (NE) and also the only fixed point of the replicator dynamics~\cite{weibull95}.
Mutual cooperation  would be socially preferable but $C$ is strongly dominated by $D$. 

\noindent In the HD game, the order of $P$ and $S$ is reversed, yielding $T > R > S > P$. Thus, in the HD
when both players defect they each get the lowest payoff. 
Players have  a strong incentive
to play $D$, which is harmful for both parties if the outcome produced happens to be $(D,D)$.
$(C,D)$ and $(D,C)$ are NE of the game in pure strategies. There is
a third equilibrium in mixed strategies which is the only dynamically stable equilibrium~\cite{weibull95}.

\noindent In the ST game, the ordering is $R > T > P > S$, which means that mutual cooperation $(C,C)$ is the best outcome,
Pareto-superior, and a NE.  The second NE, where both players defect
is less efficient but also less risky. The tension is represented by the fact that the
socially preferable coordinated equilibrium $(C,C)$ might be missed for ``fear'' that the other player
will play $D$ instead.  The third mixed-strategy NE in the game is evolutionarily unstable~\cite{weibull95}.

\noindent Finally, in the H game $R>S>T>P$ or $R>T>S>P $. In this case $C$ strongly dominates $D$ and
the trivial unique NE is $(C,C)$. The game is non-conflictual by definition and does not cause any
dilemma, it is mentioned to complete the quadrants of the parameter space.

\noindent With these conventions, in the figures that follow, the PD space is the lower right quadrant; the ST is the
lower left quadrant, and the HD is in the upper right one. Harmony is represented by the upper left
quadrant.

\begin{figure}[ht]
\begin{center}
 \includegraphics[width=7cm]{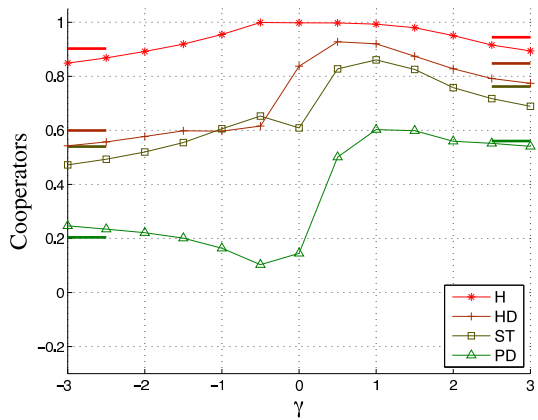}
 \caption{ (Color online) The points show, for each game, the average fraction of cooperation at steady state in the whole
 game's phase space as a
 function of $\gamma$, with weight-degree correlations $w_{ij}=(|k_i^2-k_j^2|+1)^\gamma$ in weighted Barab\'asi--Albert networks. 
 The strategy update rule is replicator dynamics.
 Averages over $50$ independent
 runs.}
 \label{gamma-rd}
\end{center}
\end{figure}


\subsection{Population Structure}
\label{pop-str}

\begin{figure*}[ht]
\begin{center}
 \includegraphics[width=17cm]{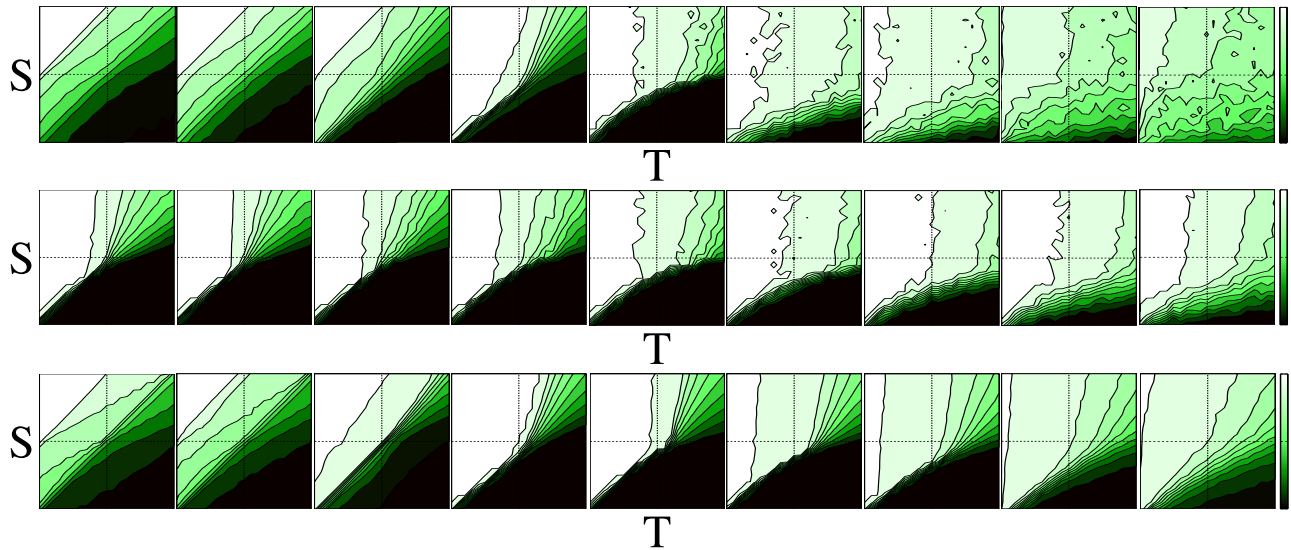}
 \caption{(Color online) Average cooperation over 50 runs at steady state on networks of size $N=2000$ and $\langle k \rangle = 8$ as a function of the parameter $\gamma$ (see text). 
 For all lines of images, $\gamma = -3, -2, -1, -0.5 , 0, 0.5, 1, 2, 3$ from left to right. The initial density of cooperators is $0.5$ in all cases. In all images the x-axis corresponds to
 $0 \le T \le 2$, and the y-axis represents the interval $-1 \le S \le 1$. Dark tones mean more defection; the color bar on the right goes from $0$ cooperation to
full cooperation at the top. Upper line: weighted BA graph; Middle line: weighted BA graph, with unweighted payoffs (see text); Bottom line:  weighted Erd\"os-R\'enyi random graphs.}
\label{phase-space}
\end{center}
\end{figure*}

\begin{figure*}[ht]
\begin{center}
 \includegraphics[width=13cm]{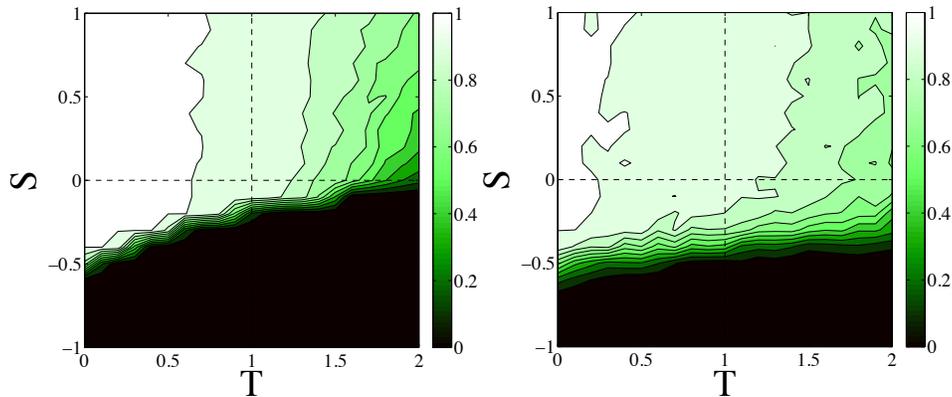}    
 \caption{(Color online) Average cooperation over 50 runs in Barab\'asi--Albert networks for $\gamma= 0.0$ on the left and $\gamma=1.0$ on the right. The initial cooperation is $0.3$ and
 strategy update is by  replicator dynamics.}
\label{coop30}
\end{center}
\end{figure*}

 \begin{figure}[ht]
\begin{center}
 \includegraphics[width=7cm]{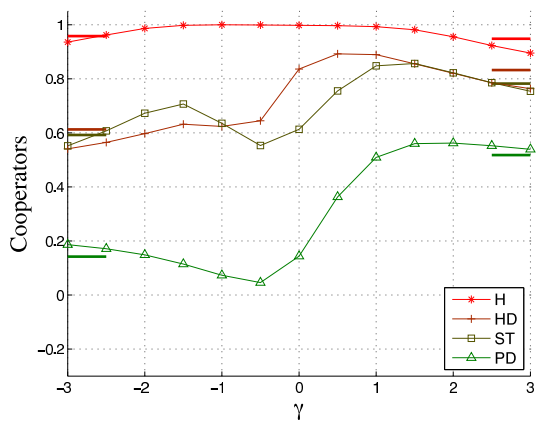}
 \caption{(Color online) The points show, for each game, the average fraction of cooperation in Barab\'asi--Albert networks 
 at steady state in the whole game's phase space as a
 function of $\gamma$ for the rule $(|k_i-k_j|+1)^\gamma$ using replicator dynamics.}
\label{diff-k}
\end{center}
\end{figure}

The population of players is a connected, weighted, undirected  graph $G(V,E)$, where the
 set of vertices $V$ represents the agents, while the set of edges  $E$ represents their symmetric interactions. The
 population size $N$ is the cardinality of $V$. The set of neighbors of an agent $i$ is defined as: $V_i =\{j \in V\: |\: \mathit{dist}(i,j)=1\}$, 
and its cardinality is the degree $k_i$ of vertex $i \in V$. The average
degree of the network is called $\langle k \rangle$. Given two arbitrary nodes $k,l \in V$, the weight of the link $\{kl\}$ is called $w_{kl}$.

\noindent For the network topology we
 use the classical Ba\-ra\-b\'asi--Albert (BA)~\cite{alb-baraba-02} networks which
are grown incrementally starting with a clique 
of $m_0$ nodes and adding a new node with $m \le m_0$ edges at each time step.
The probability that a new node will be connected to node $i$ depends on
the current degree $k_i$ of the latter: the larger $k_i$ the higher the connection probability. The model evolves into a stationary network
with power-law probability distribution for the vertex degree $P(k) \sim k^{-\alpha}$, with
$\alpha\sim 3$. 
For the simulations, we started
with a clique of $m_0=9$ nodes and, at each time step, the new incoming node has $m=4$ links. In addition, we also used
standard Erd\"os-R\'enyi random graphs~\cite{bollobas} and scale-free random graphs~\cite{Molloy}, as explained in Sect.~\ref{results}.

\subsection{Payoff Calculation and Strategy Update Rules}
\label{revision-protocols}

 \begin{figure}[ht]
\begin{center}
 \includegraphics[width=7cm]{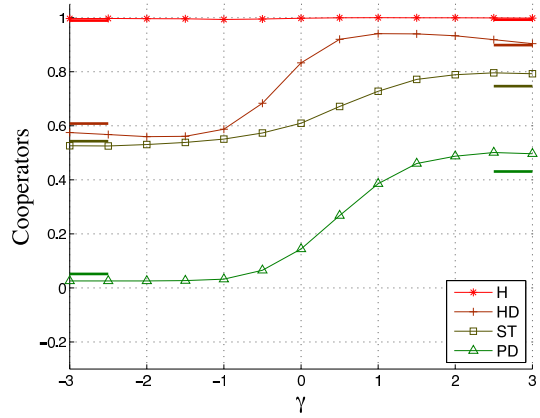}
 \caption{(Color online) Average cooperation levels for the replicator dynamics in Barab\'asi--Albert networks
 when payoff is computed according to the unweighted network (see text).}
\label{unweightedPayoff}
\end{center}
\end{figure}

We need to specify how individual's payoffs are computed and how
agents decide to revise their present strategy, taking into account that only local interactions are permitted.
\noindent Let $\sigma_i \in \{C,D\}$ be the current strategy of player $i$ and let us call $M$ the payoff matrix of the game. The quantity
\begin{equation}
\Pi_i(t) =  \sum _{j \in V_i} \sigma_i(t)\; M\; \sigma_{j}^T(t)
\label{payoffs}
\end{equation}
\noindent is the cumulated payoff collected by player $i$ at time step $t$. Since we work with weighted networks, the pairwise
payoffs $M_{ij}=\sigma_i\; M\; \sigma_{j}^T$ are multiplied by the weights $w_{ij}$ of the corresponding links before computing the
accumulated payoff $M^{'}_{ij}$ earned by $i$. This takes into account the relative importance or frequency of the relationship as represented by its weight.
Thus, the modified cumulated payoff of node $i$ at time $t$ is $ \widehat \Pi_i = \sum_{j \in v_i} \: M^{'}_{ij}$. However, if the weights are seen simply as
an expression of trust,
 the payoff is computed with the unweighted network and the weights affect only the strategy update rule, because, as the frequency of the relationship is held constant, the additional trust only influences which neighbor the player wants to imitate.

Several strategy update rules are customary in evolutionary game theory.
Here we shall describe two imitative update protocols that have been used in our simulations.

\noindent The \textit{local fitness-proportional} rule is stochastic and gives rise to replicator dynamics~\cite{hauer-doeb-2004}. 
 Player $i$'s strategy $\sigma_i$ is updated by drawing
another player $j$  from the neighborhood $V_i$ with probability proportional to the weight $w_{ij}$ of the link,
and replacing $\sigma_i$ by $\sigma_j$ with probability: 
$$
p(\sigma_i \rightarrow \sigma_j) = (\widehat{\Pi}_j - \widehat{\Pi}_i)/K,
$$
If $ \widehat{\Pi}_j >\widehat{ \Pi}_i$, and keeping the same strategy if  $\widehat{ \Pi}_j \le\widehat{ \Pi}_i$,
where $\widehat{\Pi}_j -\widehat{\Pi}_i$ is the difference of payoffs earned by $j$ and $i$ respectively.
$K=\max(s_i,s_j)[(\max(1,T)-\min(0,S)]$ ensures
proper normalization of the probability $p(\sigma_i \rightarrow \sigma_j)$, in which $s_i$ and $s_j$ are the
strenghts of nodes $i$ and $j$ respectively, the strength of a node being the sum of the weights of the edges emanating from this
node.

\noindent The second strategy update rule is the \textit{Fermi rule}~\cite{szabo}: 
$$p(\sigma_i \rightarrow \sigma_j) =\frac{1} { 1+ \exp(-\beta(\widehat{\Pi}_j - \widehat{\Pi}_i))}.$$
This gives the probability that player $i$ switches from strategy $\sigma_i$ to $\sigma_j$, where $j$ is a 
neighbor of $i$ chosen with probability proportional to the weight $w_{ij}$ of the link between them.
The parameter $\beta$ gives the amount of noise: a low $\beta$ corresponds to high probability of error and, conversely,
high $\beta$ means low error rates.

 \subsection{Simulation Parameters}
 
The networks used in all simulations are of size $N=2000$ with mean degree $\langle k \rangle=8 $.
 The $TS$-plane has been sampled with a grid step of $0.1$ and
 each value in the phase space reported in the figures is the average of $50$ independent runs, using a fresh graph 
 realization for each run.  
 
 \begin{figure*}[ht]
\begin{center}
 \includegraphics[width=13cm]{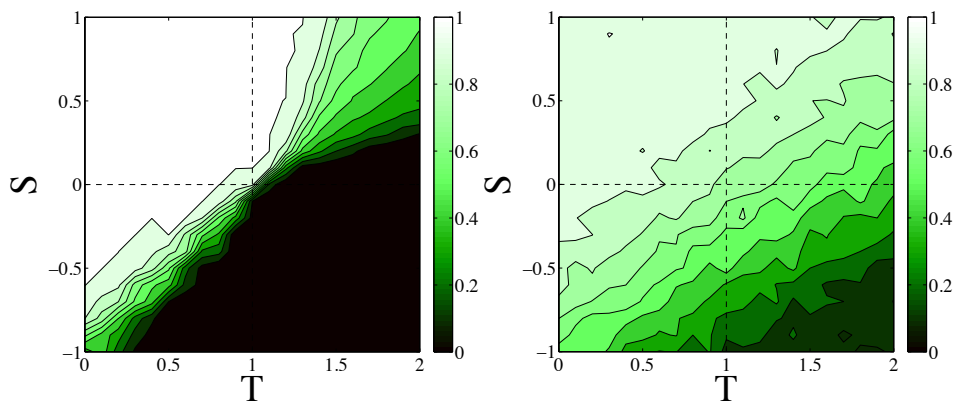}   
 \caption{(Color online) Average cooperation in Barab\'asi--Albert networks over 50 runs for $\gamma= 0$ on the left and average cooperation over 200 runs for $\gamma=1$ after 12000 synchronous time steps on the right. Both cases belong to the diffusion approximation, that is Fermi rule with $\beta= 0.01$.
}
\label{diffusion}
\end{center}
\end{figure*}
 
 The evolution proceeds by first initializing the players at the nodes of the network with one of the two strategies at random
such that each strategy has a fraction of approximately $1/2$. Other proportions have also been used.
Agents in the population are given
opportunities to revise their strategies all at the same time (synchronous updating). We have also checked that asynchronous
sequential update gives similar results.
 We let the system evolve for a period of $4000$  time steps, after any transient
 behavior has died out, and we take average cooperation values. At this point the system has reached a
 steady state in which there is little or no fluctuation.


\section{Degree-Weight Correlations}
\label{deg-weight-corr}

In a recent work Du et al.~\cite{chinois08}, based on observed degree-weight correlations in transportation networks~\cite{vespignani-aerei},
assumed that $w_{ij} \propto (k_i k_j)^\gamma$ for some
small exponent $\gamma$, where $w_{ij}$ is the weight of edge $\{ij\}$, and $k_i,k_j$ are the degrees of its end points.
The authors used 
Barab\'asi-Albert model graphs which are, among the standard model complex networks, those that are more conducive to
cooperation~\cite{santos-pach-05,santos-pach-06}. Although they only presented results for the points in the phase space belonging to the frontier segment between the PD and HD games (so-called ``weak'' PD game), here we provide the full average results for each game's parameter space as a function 
of the exponent $\gamma$ in Fig.~\ref{cadran} using replicator dynamics. The complete numerical results for the whole phase space are to be found in~\cite{weighted-physicaA}.

For $\gamma \ge 0$, there is indeed a small but non-negligible increase in cooperation around $\gamma=0.0$ to $0.5$. However, the problem is that the
assumed degree-weight correlation is typical of transportation networks but has not been observed in social
 networks~\cite{vespignani-aerei,mobile-nets1,mobile-nets2}. The reason could be that while in transportation networks there are
fluxes that must respect local conservation~\cite{mobile-nets2}, social networks have a more local nature.  
In the degree-weight correlation above the coordinate $k_i k_j$ is the joint contribution; a complementary view would be to use 
the difference between $k_{i}$ and $k_{j}$ which amounts to a local degree comparison. 
Thus, as there is no correlation between the form $k_i k_j$ and real weights, we take its perpendicular 
counterpart $|k_i^2-k_j^2|$ in the degree-weight correlation, which contains the same information as the absolute value of the difference
of degrees. 
Taking these considerations into account we have:

\begin{equation}
w_{ij} \propto (|k_i^2 -k_j^2|+1)^\gamma
\label{corr}
\end{equation}

The exponent $\gamma$ allows us to explore the weight-degree correlation from a more disassortative case ($\gamma > 0$) to an assortative case ($\gamma < 0$), passing through the unweighted case ($\gamma = 0$). We added a strictly positive constant to the difference of square degrees to avoid division by $0$ for negative $\gamma$. 

It is important to point out that we do not mean to imply that this kind of correlation is present in actual
social networks as we do not have the empirical data analysis needed to show that this is the case. Indeed, to our knowledge, there are not 
enough publicly available reliable data on weighted social networks to test it out empirically. 
As a consequence, our results will refer to model weighted networks, not
to real ones, but this has  been the case for most studies of evolutionary games on unweighted networks as well (~\cite{santos-pach-05,santos-pach-06,szabo,anxo1}
and references therein).

\begin{figure}[ht]
\begin{center}
 \includegraphics[width=7cm]{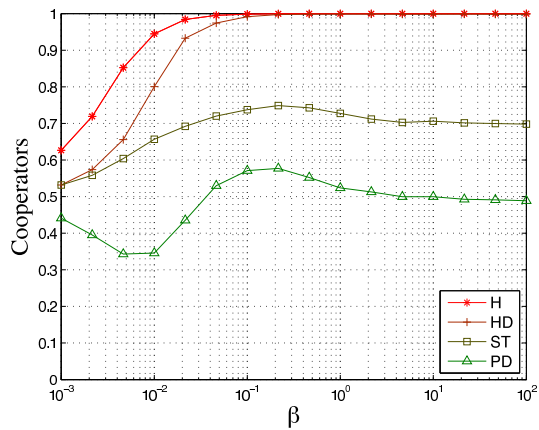} 
 \caption{(Color online) Average cooperation in all games on Barab\'asi--Albert networks over 100 runs for $\gamma= 1$ after 4000 synchronous time steps with the Fermi rule as a function of $\beta$. 
 }
\label{diffusion2}
\end{center}
\end{figure}

 \section{Results}
 \label{results}
The average cooperation figures for BA networks, in which weights are set according to the degree-weight correlation described
by Eq.~\ref{corr}, are shown in Fig.~\ref{gamma-rd} as a function of the exponent $\gamma$. The strategy update rule is replicator dynamics.
The increase in cooperation is notable in the positive $\gamma$ range, especially in the hardest PD game, well beyond what is found in unweighted BA networks, as can be seen in Figs.~\ref{phase-space} by comparing the 5th image in the top panel, which 
corresponds to an unweighted network ($\gamma=0$), with the next images in the same panel. Moreover, it is seen
that the HD and ST games also benefit in this range. Taking into account that BA scale-free networks have been found to be among the most effective to date in terms of amplifying cooperation, these findings seems to be very interesting. 
In Fig.~\ref{phase-space} it can be seen that, in contrast to the positive $\gamma$ case, for negative $\gamma$ there is no cooperation gain
whatsoever. A qualitative explanation of this result is the following. When $\gamma < 0$, Eq.~\ref{corr} attributes small weights to links between
nodes that have widely different degrees. Because of the way in which weights are used to choose neighbors and to compute payoffs
in the game dynamics, this in turn will prevent hubs from playing their role in propagating cooperation. The situation for $\gamma=-0.5$ becomes
qualitatively similar to that of a random graph in which the degree distribution is more homogeneous, and it is well known that random graphs
are not conducive to cooperation~\cite{anxo1}. When $\gamma$ decreases further and approaches $-3$, in practice the network becomes more 
and more segmented into small components that have very weak links between them. 
 \begin{figure}[ht]
\begin{center}
 \includegraphics[width=7cm]{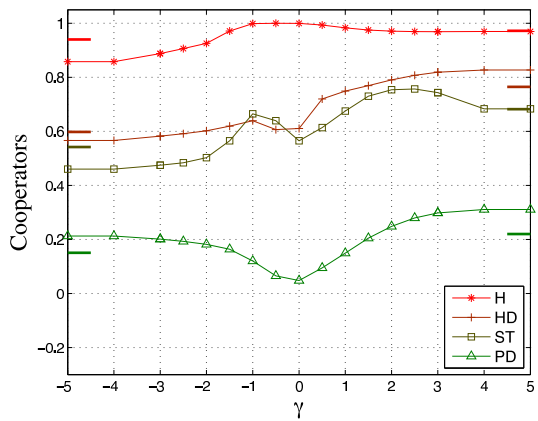}
 \caption{(Color online) The points show, for each game, the average fraction of cooperation at steady state in the whole
 game's phase space as a
 function of $\gamma$ for  Erd\"{o}s-R\'{e}nyi random graphs. Strategy update is by replicator dynamics and the
 points are averages over $50$ independent runs.}
\label{gamma-rd-ra}
\end{center}
\end{figure}


As expected, the initial conditions influence final cooperation frequencies. As an example, we display the phase space at steady state starting from $30$ percent cooperation in Fig.~\ref{coop30} for $\gamma = 0$ and $1$. The average cooperation is less, especially for the SH and PD games; however, 
cooperation levels are still significantly higher with the weights.

The rule $(|k_i^2-k_j^2|+1)^\gamma$ uses the square of the degrees. We also tested the simpler alternative rule $(|k_i-k_j|+1)^\gamma$ on a Barabasi-Albert network. We obtained the same behavior, except for a shift,  see Fig.~\ref{diff-k}.

Until now we used the weights as representing a time of interaction, and thus payoffs were the product of the weight times the numerical 
payoffs as explained in Sect.~\ref{revision-protocols}. We can also consider them simply as an expression of trust. In this case 
the weights only influence the choice of a neighbor to imitate,  but the payoffs are just the ordinary payoffs, Eq.~\ref{payoffs}, without weight multiplication. For the PD, the assortative weights ($\gamma<0$) imply close to zero cooperation, and the disassortative weights tend to one half cooperation. We show the mean cooperation for all games in Fig.~\ref{unweightedPayoff}, and the entire game's space in the middle panel of Fig.~\ref{phase-space}, using the rule $(|k_i^2-k_j^2|+1)^\gamma$. We observe that all the games are positively influenced
for $\gamma > 0$ which indicates that simply imitating the strategy of agents with which the link is stronger has a favorable effect on cooperation.

The Fermi dynamics (see Sect.~\ref{revision-protocols}) is a flexible imitation protocol. For $\beta \simeq 10$ it approaches the replicator dynamics
results in unweighted networks~\cite{anxo1}. However, when $\beta \ll 1.0$  the Fermi dynamics leads to the diffusion approximation~\cite{szabo}.
Here, this approximation means that  when a player selects a strategy, after having selected a neighbor, in the majority of cases he chooses  the strategy at random and 
only rarely he selects it proportionally to difference of payoffs. For unweighted BA networks at steady state there is no cooperation increase and defection
prevails in the PD (see left image of
Fig.~\ref{diffusion} and~\cite{anxo1}). 
However, on the weighted network with $\gamma=1.0$, cooperation remains high, although lower than the replicator dynamics
case (Fig.~\ref{diffusion}, image on right). 
We performed $200$ runs instead of $50$ because the system is noisier, and each run lasted for $12000$ instead of $4000$ time steps,
since approaching steady state is slower. In Fig.~\ref{diffusion2} we display the cooperation levels as a function of $\beta$ for the weighted graph with $\gamma=1$ after $4000$ synchronous time steps and $50$ repetitions. This seems to confirm the robustness of cooperation, on this kind of topology, against noise in the choice of a neighbor. For $\beta\le10^{-2}$ the dynamics is probably not at steady state.
Steady state might be difficult to attain because the dynamic becomes very slow as $\beta$ approaches $0$. However, the difference from the BA is very interesting (See also~\cite{anxo1});
 it shows that the model is applicable to a broader range of real situations, where the behavior of players is subject to errors of various kinds. 

   \begin{figure}[ht]
\begin{center}
 \includegraphics[width=7cm]{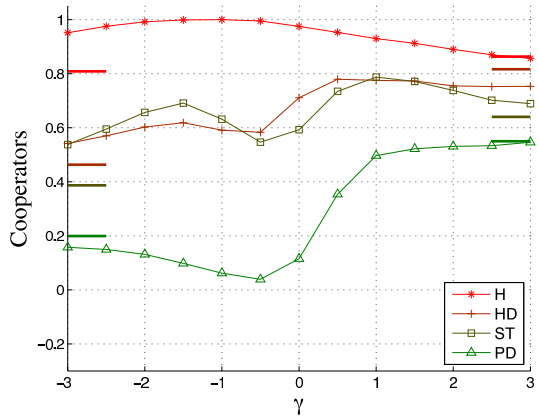}
 \caption{(Color online) Average cooperation in random scale-free graphs obtained with the configuration 
 model with an exponent $\alpha=-3$ of the power law degree distribution. Strategy update rule is replicator dynamics.
 Averages over $50$ independent runs.}
\label{ConfModel}
\end{center}
\end{figure}

In order to see how cooperation is affected when degree heterogeneity is low, we use  Erd\"{o}s-R\'{e}nyi random graphs and then set the weights according to our rule. The Barabasi-Albert graph gives drastically better results but there is an improvement in weighted random graphs compared to the 
unweighted case: see Fig.~\ref{gamma-rd-ra} and the bottom panel of Fig.~\ref{phase-space}. Indeed, while cooperation in the PD is almost absent on random graphs, with our weighting scheme, cooperation increases  to levels similar to those found in unweighted BA networks, as seen by comparing Fig.~\ref{gamma-rd-ra} for $\gamma=1$ with
Fig.~\ref{gamma-rd} for $\gamma=0$.

It is worth noting that Barab\'asi--Albert graphs are only a particular class of scale-free growing networks. Due to their construction, they possess historical
correlations between early hub nodes~\cite{newman-book}. In other words, hubs tend to be interconnected among themselves and it
transpires that this feature is favorable for cooperation~\cite{santos-pach-05,anxo1}. Thus, we also tested scale-free networks
that have no degree correlation, such as the ``configuration model''~\cite{Molloy,newman-book} which yields random scale-free
graphs if built from the right degree sequence. We used  the configuration model with an exponent $\alpha= -3$ of the power-law degree distribution. The results, shown in Fig.~\ref{ConfModel},
 are qualitatively 
similar to those obtained on BA networks (Fig.~\ref{gamma-rd}), except  for a small drop in cooperation near $\gamma=1$. The largest differences
are observed for the ST game, but they are still small in absolute terms. 
The result on random scale-free weighted networks is interesting and
confirms the role of weights in the establishment of asymptotic high levels of cooperation in our model networks, in spite of the fact that
these networks are less conducive to cooperation than BA networks~\cite{santos-pach-06,poncela-int-jbc}.

\section{Cooperative Unweighted Graphs from Weighted Ones}

 \begin{figure*}[ht]
\begin{center}
 \includegraphics[width=13cm]{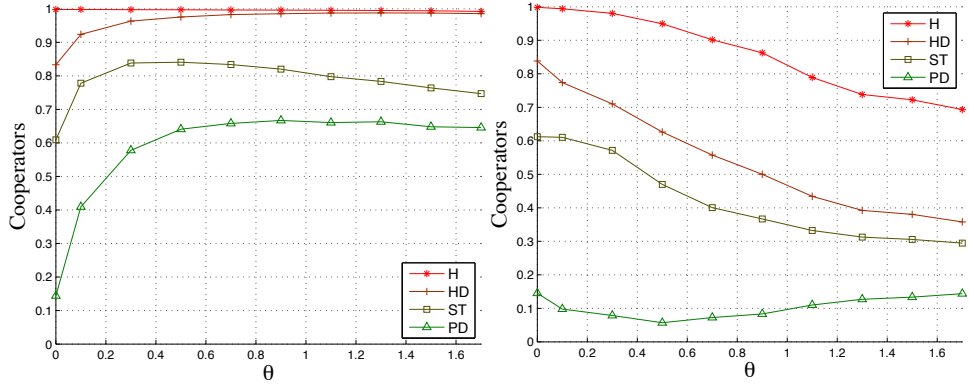} 
 \caption{(Color online) Average cooperation  
 for the filtered unweighted graphs obtained from Barab\'asi--Albert networks with $\gamma=1.0$ (left) and $\gamma=-1.0$ (right), as a function of the parameter $\theta$ (see text). 
 Values are averages over $50$ runs, and strategy update is by replicator dynamics.}
\label{beta}
\end{center}
\end{figure*}

Understanding game dynamics may be simpler on unweighted graphs than on weighted graphs. 
In order to get an intuitive idea of the mechanism seen on weighted networks, we constructed two unweighted models from the weighted one as described below. 

The first model is a rather raw but revealing way of extracting
an unweighted structure from the weighted graph by filtering out edges with low enough weights. The resulting topology should be of help in understanding the
origins of the large promotion of cooperation we found on our particular class of weighted correlated networks.
Starting with a BA graph, we set the weights according to the form $(|k_i^2-k_j^2|+1)^\gamma$, 
discard the edges with a weight inferior to a threshold $\theta$, and finally set all weights of the remaining links to one. 
 Every edge with weight
  $w < \theta \: \langle w \rangle$, 
 $\theta$=\{0.0,0.1,0.3,...,1.9\}, is discarded. During the edge filtering process some vertices may become isolated. The problem is that their strategy cannot change after having been assigned randomly at the beginning. In the weighted network these nodes interact rarely since the frequency of interaction depends on
 the link weight.  For the sake of simplicity, our choice has been to discard them in the simulations, otherwise the simulation results would be biased
 owing to the isolated nodes with unchanging strategy.
 
 \begin{figure*}[ht]
\begin{center}
 \includegraphics[width=13cm]{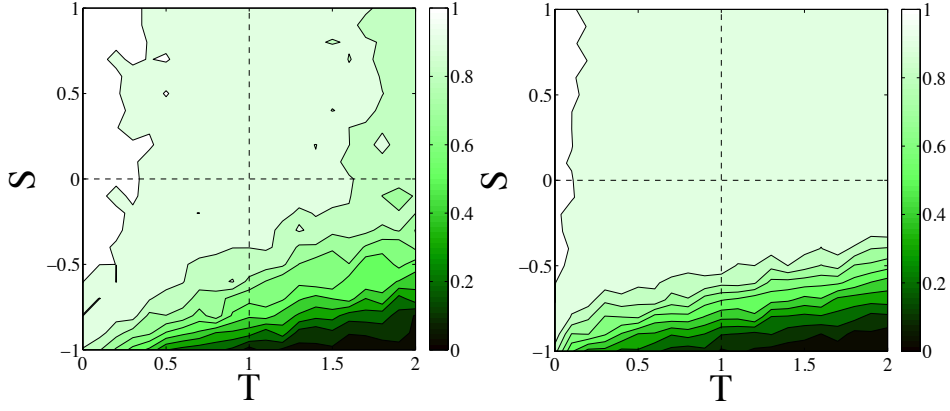} 
 \caption{(Color online) Left : Cooperation phase space on Barab\'asi--Albert networks for the weighted case with $\gamma = 1$ and the correlation 
 function $(|k_i^2-k_j^2|+1)^\gamma$. Right : Corresponding cooperation phase space for the filtered unweighted network with
 $\gamma=1$ and  $\theta=0.7$.
 Averages over $50$ runs using replicator dynamics.}
\label{beta1}
\end{center}
\end{figure*}

 The mean cooperation on the filtered graphs as a function of $\theta$ is shown in the left image of Fig.~\ref{beta}. The mean cooperation level in all the games
 increases with the threshold up to a certain point and then stays approximately constant. The cooperation increase is particularly notable
 for PD in which one goes from $0.15$ up to about $0.65$, i.e. a four-fold increase.
 The right image of Fig.~\ref{beta1} shows the average cooperation in the whole S-T plane for $\gamma=1$ and $\theta=0.7$. It is apparent that the high degree
 of cooperation that was present in the original weighted graph, which is reported in the left image of Fig.~\ref{beta1} is fully maintained apart from some small differences due to the details of the dynamics and slightly different network sizes. This means that the filtered unweighted graph provides at least 
 as much cooperation as the original weighted one and, therefore, is in some non-rigorous sense, equivalent to it. For higher values of $\theta$, not
 only in the average but also for the whole game space, the
 trend remains very similar for the PD; in the ST there is some small loss of cooperation, while the HD gains a little bit (the corresponding figures
 are not shown to save space).
  
 In the right image of Fig.~\ref{beta} we show the mean cooperation values that are reached when the links are filtered in the same
 way but with $\gamma=-1$. Since the weights are now the reciprocals of the original ones, this roughly amounts to cutting the strongest links 
 with respect to the case  $\gamma=1$. The result is a generalized loss of cooperation in the average, a fact that shows the importance of
 strong links. The fact that cooperation goes below $1$ in the Harmony game is due to graph fragmentation.

\begin{figure*}[ht]
\begin{center}
\begin{tabular} {cccccccc} 
 \includegraphics[width=4.6cm,height=4.6cm]{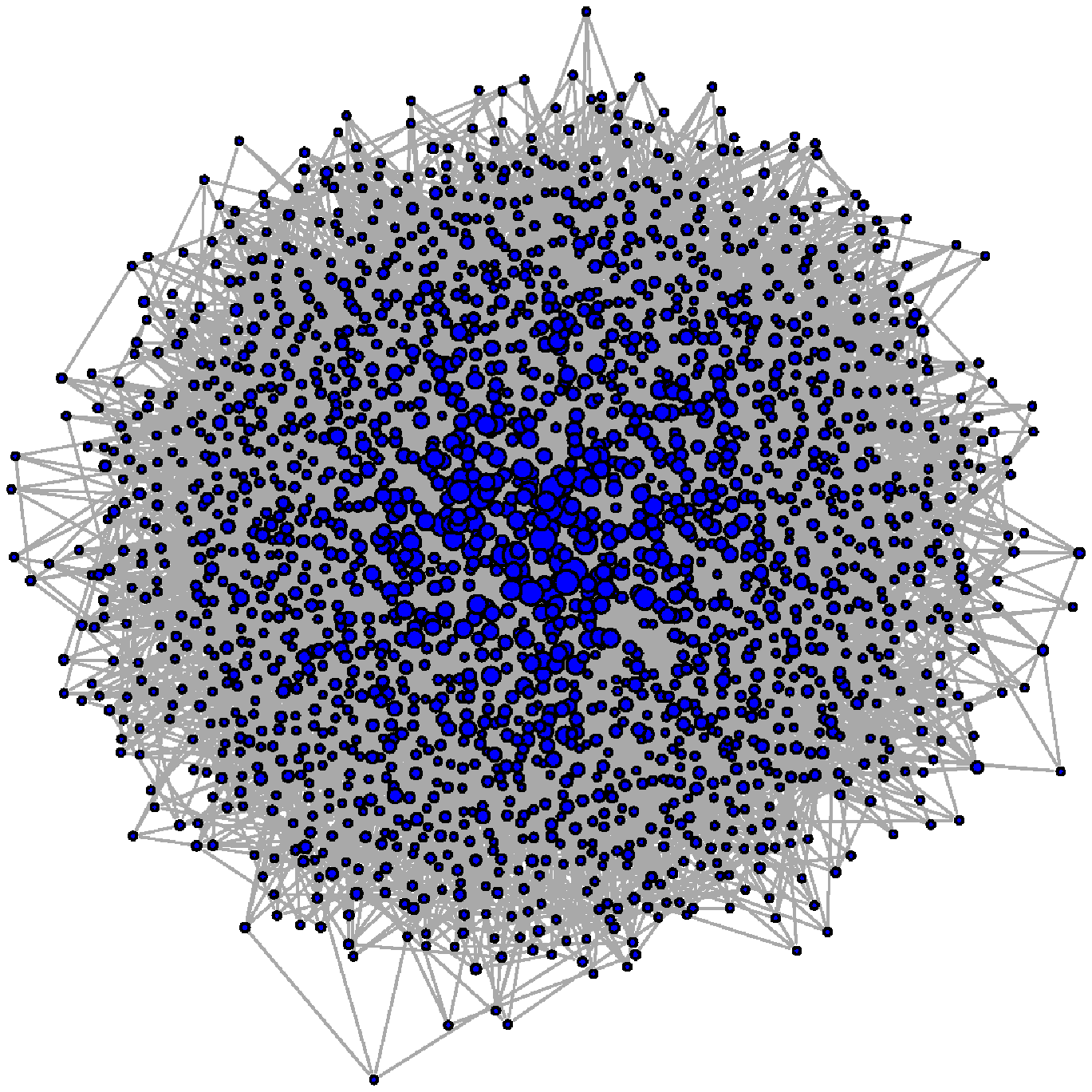}  &  
 \includegraphics[width=4.6cm,height=4.6cm]{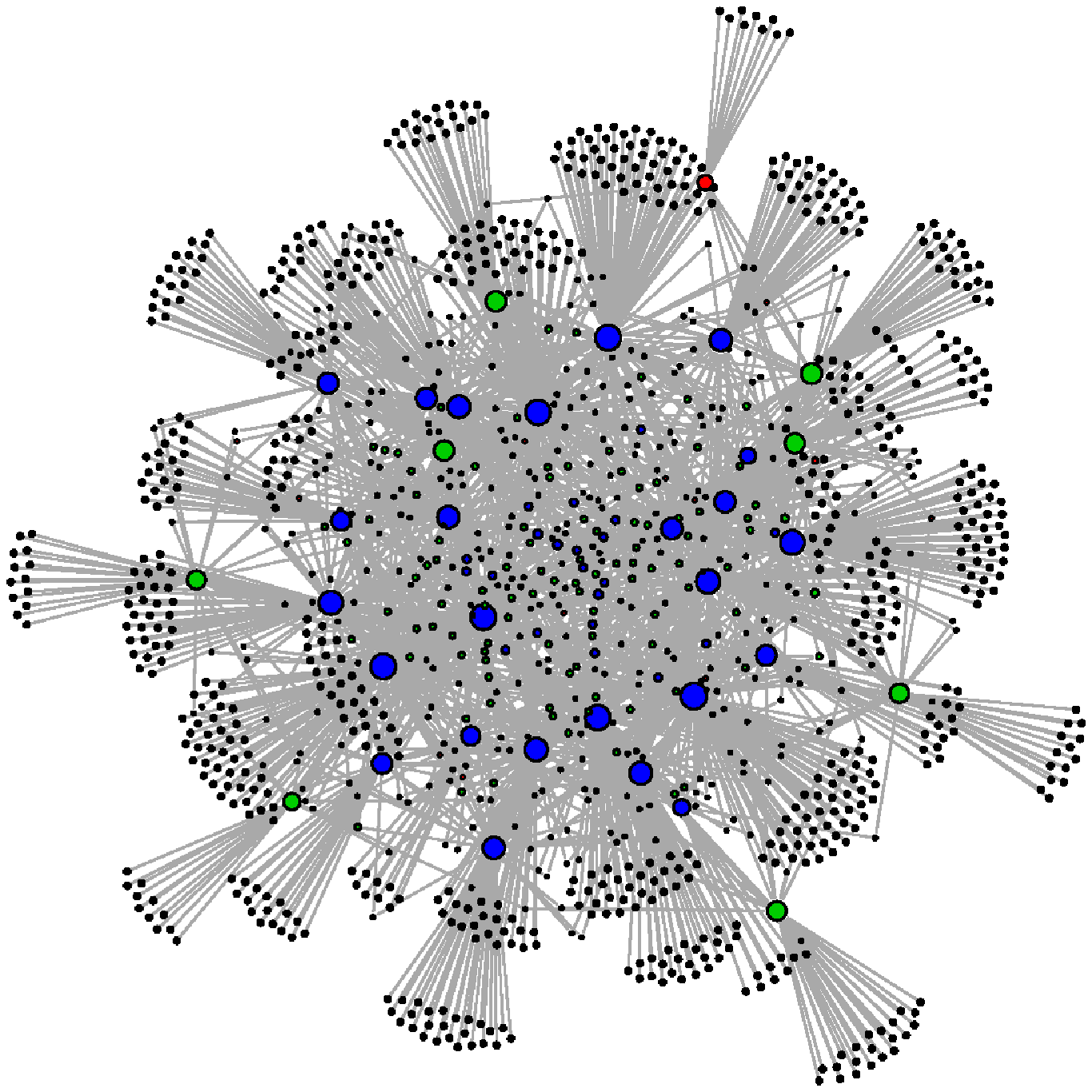}  &  
  \includegraphics[width=4.6cm,height=4.6cm]{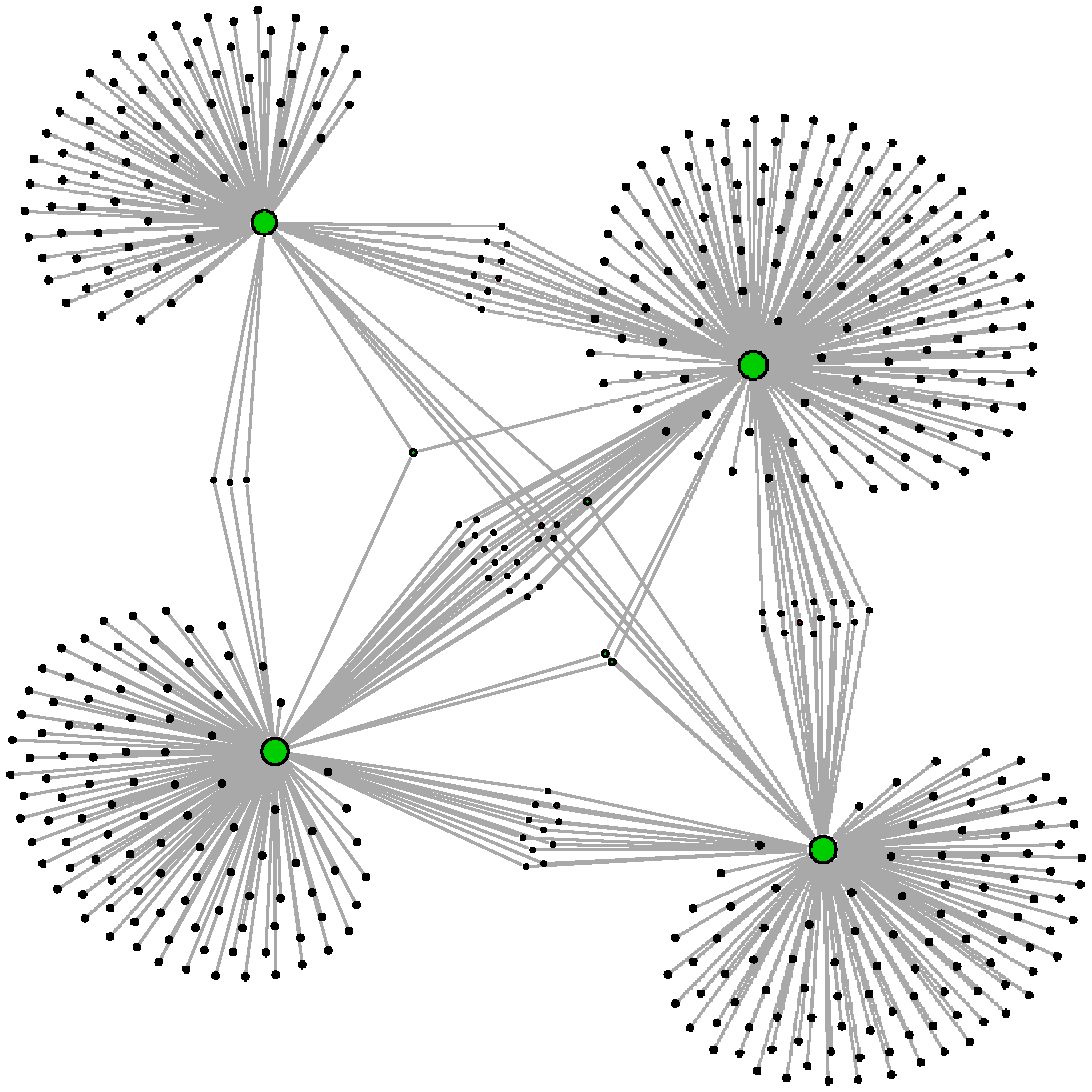}  &  
\end{tabular}
 \caption{(Color online) Typical filtered graph plots as a function of $\gamma$, with Fruchtermann-Reingold layout, coloration of cores, and the size of a vertex  $\propto log(k_i)$. 
 From left to right: $\gamma = 0.0, 1.0, 3.0$; $\theta=0.7$ for all images. Only the giant connected components are shown. }
\label{unweightedGraphs}
\end{center}
\end{figure*}

 \begin{figure}[ht]
\begin{center}
 \includegraphics[width=7cm]{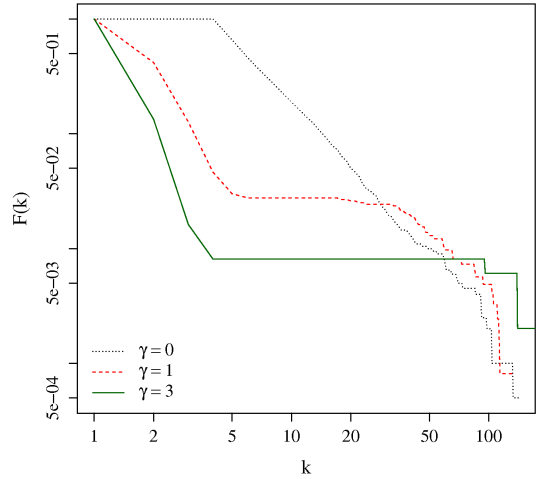}   
 \caption{(Color online) Cumulated degree distribution of the filtered and unfiltered networks in log-log scales. 
 dotted gray line, dashed red line, green thick line stand, respectively, for $\gamma = 0.0, 1.0, 3.0$. The original
 unfiltered graph has been obtained through the Barab\'asi--Albert construction.}
\label{DegreeDiscard}
\end{center}
\end{figure}

 We display some particular but representative graphs obtained through the filtering process in Fig.~\ref{unweightedGraphs}. 
 The topologies show hubs connected to many low degree vertices and
 the hubs are possibly connected directly or through some intermediate vertices with low degree, while the low degree vertices are not connected between them.
 This effect increases with increasing $\gamma$ and can be seen
 explicitly in Fig.~\ref{unweightedGraphs}. The empirical degree distribution curves of the BA graphs before and after 
 transformation are shown  in Fig.~\ref{DegreeDiscard}. We  observe that there are less nodes with degree between $\simeq 5$ and $\simeq 50$, thus separating vertices 
 in two sets, large degree and low degree vertices, an almost bipartite network. In fact, the Newman's coefficient of assortativity~\cite{newman-book} is of the order of $\sim-0.6$,  for the filtered BA graph 
 with $\theta=0.7, \gamma=1$, and  of the order of $\sim-0.9$ for $\theta=0.7, \gamma=3$, while it is $\sim-0.05$ for the original graph, which indicates that the graph indeed becomes
  more bipartite as there are few links between nodes of similar degree and $\gamma$ increases. Another class of networks in which one
  finds this almost bipartite degree distributions have been generated by Poncela et al.~\cite{moreno-evol} in a completely
  unrelated way. They use a dynamical process in which new players
  attach to existing nodes at random, or preferentially  to those  that have been successful in the past. Their graphs are highly
  cooperative in the dynamical regime, but when used as static graphs cooperation is much lower. 
   In another study, Rong et al.~\cite{Rong} find that when a network becomes assortative by degree, 
the large-degree vertices tend to interconnect to each other closely, which destroys the sustainability among cooperators and promotes the invasion of defectors, whereas in disassortative networks, the isolation among hubs protects the cooperative hubs in holding onto their initial strategies to avoid extinction.
This study, although it is different from ours since it does not start from weighted networks, also shows the role of degree disassortativity
in influencing cooperation.


The above model gave us some general ideas about the possible structure of highly cooperative unweighted networks. Now we shall try to
better understand the microscopical foundations of the phenomena by studying
a small graph containing the minimal features required to parallel the behavior on  weighted networks. 

In Roca et al.~\cite{anxo1} the authors give an intuitive explanation for the cooperation induced by degree heterogeneity in
unweighted BA networks. 
G\'omez-Garde{\~{n}}es et al.~\cite{moreno-coop-PRL} provide a deeper analysis of the origins of cooperation in the PD case
but here the simple argument of~\cite{anxo1} will be sufficient.
In their view, the hubs drive the population, fully or partially, to cooperation in HD games. The reason is the following: the hubs get more payoff, thus a defector hub triggers the change of his neighbors to defection, while a cooperator hub triggers it to cooperation, favoring cooperation since cooperative hubs
surrounded by cooperators get a higher payoff.
In our weighted case the topology of the network leads to cooperation not only in the HD game, but also in the PD game. In this case the idea of hubs driving the population to cooperation is very relevant, because the new topology amplifies this effect.

\begin{figure*}[ht]
\begin{center}
  \includegraphics[width=13cm]{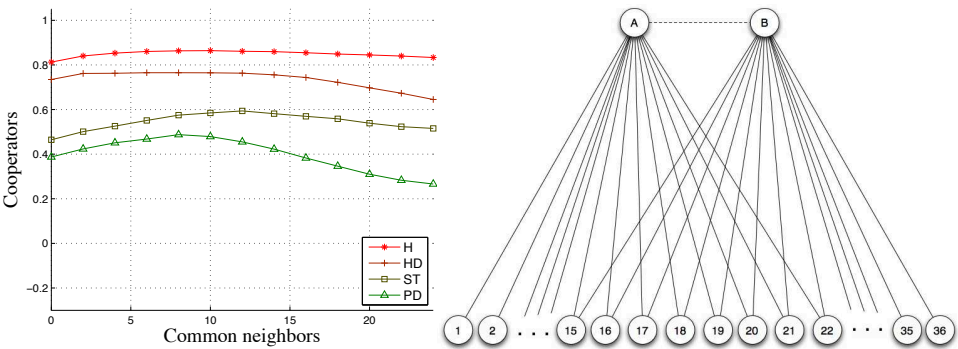}
 \caption{(Color online) Average cooperation over 1000 repetitions of 500 time steps on the bipartite graph shown on the right as a function of the number of common neighbors to A and B. Replicator dynamics is used to update agents' strategy. The initial strategy distribution is random. The graph displayed has 8 common neighbors.}
\label{bipartit1}
\end{center}
\end{figure*}
\begin{figure*}[ht]
\begin{center}
 \includegraphics[width=13cm]{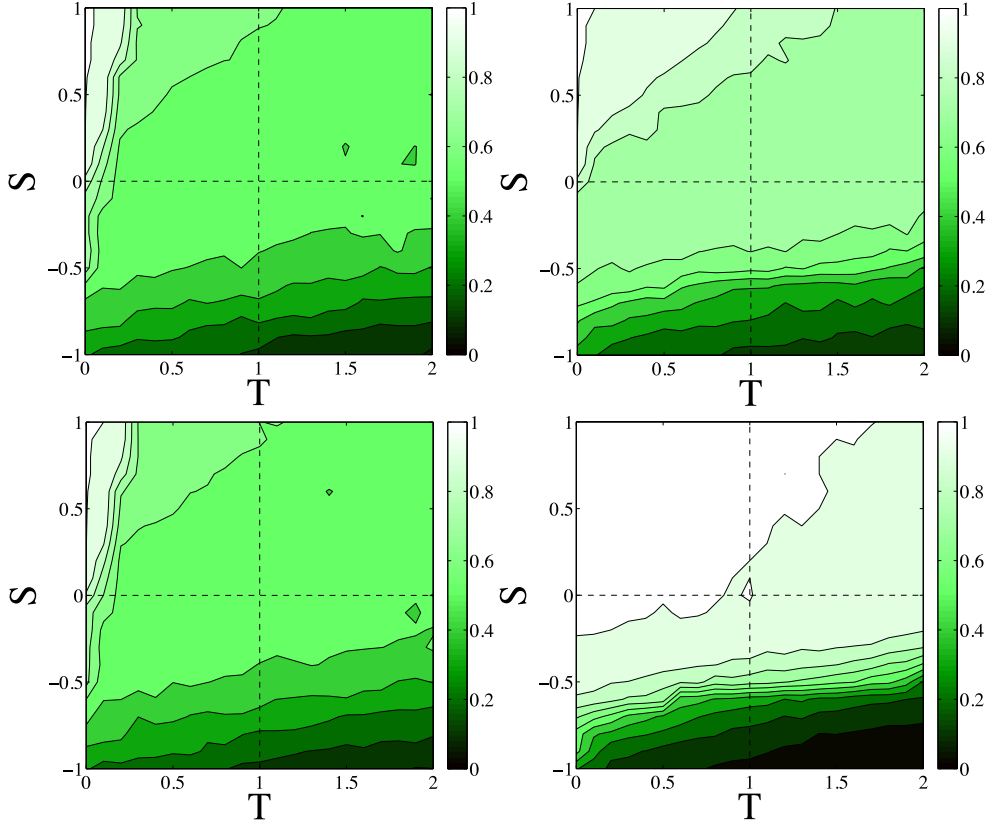}  
 \caption{(Color online) Average cooperation over 1000 repetitions of 500 time steps for 8 common neighbors without a link between A and B (left), and with a link between A and B (right) (see Fig.~\ref{bipartit1}). Strategy update uses the replicator dynamics rule. Upper row: The initial strategy distribution is random; Lower row: The initial strategy distribution is random except for the hubs, one is set cooperator, the other defector.}
\label{bipartit2}
\end{center}
\end{figure*}

According to Eq.~\ref{corr} for $\gamma > 0$, the smaller the difference between the degrees of two connected nodes, the smaller
the weight of the corresponding link. Thus, two connected hubs will have a relatively low link weight, while a link between a hub
and a low-degree vertex will be given a high weight. Also, links between two low-degree nodes  will make the
corresponding link weight almost negligible, a fact that we have exploited in the filtered graphs above. With these ideas in mind,
let us consider  the graph in the right part of Fig.~\ref{bipartit1} where two hubs are connected to $36$ vertices having degree one or two, depending on whether they are connected to one hub or to both. We will consider two extreme cases: either the two hubs are connected, or they aren't.
We obtain a bipartite graph when the hubs are not connected, and an almost bipartite one when there is a link between them.
 The number of common neighbors of A and B is $8$ in the figure,
 so as to be in a region where cooperation is high, as shown in the left image of Fig.~\ref{bipartit1} when $A$ and $B$ are connected. 
 These data have been obtained by numerically simulating the games' evolution on the graph of Fig.~\ref{bipartit1} with replicator
 dynamics and varying the number of common neighbors from $0$ to $24$.
 The upper row of Fig.~\ref{bipartit2} shows the games' phase space with or without an edge between the two hubs for the graph of Fig.~\ref{bipartit1} 
 starting from a random initial distribution, while the lower row of Fig.~\ref{bipartit2} shows the same plots with the following initial distribution: A is cooperator, B is defector, random strategies are assigned to the other vertices. The plots show that the model in which the hubs are connected  induces more cooperation in all games. When
hubs are not connected (the left images) results are very similar in the upper and lower images because propagation via the hubs is no longer possible. Now, we shall try to explain the mechanism at work referring to the graph in Fig.~\ref{bipartit1}.

If node A is a cooperator and node B is a defector, by Roca's et al. arguments~\cite{anxo1}, A will have an advantage and could spread
cooperation to B. However, if A and B are not directly connected, strategy migration will have to go through a low degree
node and this will make further progress more difficult because B is a hub and is likely to collect a higher payoff. On the
other hand, when there is a link between A and B, B could imitate cooperator A at some point. To spread cooperation
further now cooperator B needs cooperator neighbors in order to get a sufficiently high payoff.  B could obtain this from
common neighbors with A, which are more likely to be cooperators than the neighbors that are only connected to it and
not to A. Therefore, having a direct connection between A and B is crucial for the dynamics to lead to high degrees of
cooperation. This can be seen in the numerical results shown in Fig.~\ref{bipartit2} for the case in which A has a
link to B (right part) and to the opposite case (left part). A certain number of shared neighbors between A and B seems
to play an important role. Without them, there would be segregation of strategies, reflected in the average cooperation level  in  
Fig.~\ref{bipartit1} (left) for $0$ or a low number of common neighbors for the HD and the PD. As this number increases, however, cooperation may
propagate from hub to hub. When the number of common neighbors becomes too high, around $13$, a cooperator hub is no longer favored as now his
cooperator and defector neighbors will be about the same number.

It is worth noting that Perc et al.~\cite{Perc-et-al} with an apparently unrelated grid population model in which a fraction of players is more influential
in the sense that they can transmit their strategy to others more easily, also obtain a high amount of cooperation
in the weak PD game. For this to happen, it is also needed that, with a small probability a player can randomly link temporarily to distant
sites in the lattice. When both features are present cooperation is boosted.

\section{Discussion and Conclusions}

Although weighted networks are closer to reality, evolutionary games on complex networks have been essentially studied on unweighted
networks until now, both for simplicity as well as because weights in social networks are notoriously difficult to assess.
In this article we introduced a new degree-weight correlation form in a model  derived from three standard unweighted networks classes: BA graphs, the
configuration model, and Erd\"os-R\'enyi random graphs.  We studied the evolutionary game dynamics of standard two-person, two-strategies games
 on these classes of networks for which accurate results exist in the unweighted cases.
A weight-degree correlation of the form $(k_ik_j)^{\gamma}$ had already been studied~\cite{chinois08,weighted-physicaA}, giving a small gain in cooperation for a restricted
range of $\gamma$ on BA networks. However, it has  empirically been found that real social weights are not correlated to $k_ik_j$~\cite{vespignani-aerei,mobile-nets2}. 
Thus in this work we used the perpendicular coordinate $|k_i^2-k_j^2|$ in the weight-degree correlation $(|k_i^2-k_j^2|+1)^\gamma$ to study the proportion of cooperators at steady state on static weighted networks, and varying the weights from a more assortative case to a more disassortative case passing through the unweighted case. 
As $\gamma$ increases toward disassortative weights, cooperation dramatically increases until a maximum of $0.6$ averaged in the whole PD game phase space.
And results are equally good for the HD and ST games, both for the replicator dynamics as well as for the Fermi rule. A small value of $\gamma > 0$ such as $0.5$ or one is sufficient to induce
unprecedented amounts of cooperation, and this remains true also for the linear correlation form $(|k_i-k_j| + 1)^{\gamma}$.
The outcome at steady state depends on the initial conditions: starting from lower cooperation fractions induce less cooperation at steady state, which is also the case in the unweighted networks~\cite{anxo1}.
The best results were obtained for BA networks or random scale-free graphs, but even with Erdos-R\'enyi random graphs, which are known to induce little or no cooperation
 in the unweighted case~\cite{anxo1}, results are better, especially for the PD in which the amount of cooperation 
 is comparable with the corresponding result for an unweighted BA graph. On the other hand, the assortative weights ($\gamma<0$) 
 are in general slightly detrimental to cooperation, except for the PD game where the increase is small anyway.
 Summing up, we can say that the main result in this part is that payoff-proportional imitation induces cooperation 
in a large part of the Prisoner's Dilemma with the given weights and $\gamma>0$ on heterogeneous networks. The other games are positively affected as well.

In the second part of the present work we provided qualitative arguments for the emergence of large increases of cooperation in 
unweighted networks derived from the weighted ones. To this end, we 
proposed to filter out edges with low weight in the original network and to set all the remaining weights to 1.
 The cooperation level in those unweighted graphs, after removing isolated vertices, is similar or even higher than in the weighted networks. The features observed are: a bipolar distribution of degree (low degree and high degree vertices), degree disassortativity, low degree vertices are connected only to hubs while some connections between hubs may still exist for moderate $\gamma$, thus a nearly bipartite graph. Then, keeping only the necessary features, we constructed a small network which helps to intuitively understand the origins of the large increase in cooperation. We found that while low degree vertices surrounding hubs advantage cooperators, common neighbors between hubs and direct links between hubs are important for the propagation of cooperation in the network.

A general consideration on the results we obtained is the following. We do not know whether the weighted networks, or the unweighted
ones derived from them
do really exist, at least in an approximate form. Ours is only an abstract model and weighted network data to perform empirical
analyses are difficult to find.
Nevertheless, in the future it would be interesting to study empirical degree-weight correlations in real-life
network. If correlations of the type assumed in this work with $\gamma>0$ could be found, then the corresponding networks would likely be
favorable for cooperation in evolutionary games.
Also, considering dynamical network evolution linked to a game~\cite{mini-rev}, cooperative network similar to ours could emerge at
steady state, because they maximize the number of satisfied players.   Nevertheless, naturally occurring social networks do have
a large amount of clustering, which is not the case in our filtered unweighted networks, but could be obtained on the weighted ones. 
Finally, the highly hierarchical cooperative unweighted structures found in this work could be created on purpose, if cooperation is the objective. This
could be especially useful when the network nodes are not only persons but also organizations or institutions that happen to interact
according to game rules similar to those used in the simulations.

\section*{Acknowledgments} We thank our colleague Alberto Antonioni for carefully reading the manuscript and for his insightful
suggestions. We also thank the anonymous reviewers for their useful and constructive comments.


\end{document}